\documentclass[twocolumn]{aastex63}

\usepackage{amssymb,amsmath,graphicx,multirow,float}

\shorttitle{Short-duration GRB Rates}
\shortauthors{Dimitrova et al.}

\begin{document}
\title{Predicting Short-duration GRB Rates in the Advanced LIGO Volume}

\author[0000-0002-4319-1615]{Tzvetelina A. Dimitrova}
\affiliation{Earth and Space Exploration, Arizona State University, P.O. Box 871404, Tempe, AZ 85287-1404, USA}

\author[0000-0002-9110-6673]{Nathaniel R. Butler}
\affiliation{Earth and Space Exploration, Arizona State University, P.O. Box 871404, Tempe, AZ 85287-1404, USA}

\author{Srihari Ravi}
\affiliation{Earth and Space Exploration, Arizona State University, P.O. Box 871404, Tempe, AZ 85287-1404, USA}

\begin{abstract}
Starting with models for the compact object merger event rate, the short-duration Gamma-ray Burst (sGRB) luminosity function, and the \textit{Swift}/BAT detector, we calculate the observed \textit{Swift} sGRB rate and its uncertainty. Our probabilistic sGRB world model reproduces the observed number distributions in redshift and flux for 123 \textit{Swift}/BAT detected sGRBs and can be used to predict joint sGRB/LIGO detection rates. We discuss the dependence of the rate predictions on the model parameters and explore how they vary with increasing experimental sensitivity. In particular, the number of bursts in the LIGO volume depends strongly on the parameters that govern sGRB beaming. Our results suggest that nearby sGRBs should be observed to have broader jets on average ($\theta_{\rm jet}\gtrsim 30$ degrees), as compared to the narrowly-beamed appearance of cosmological sGRBs due to detection selection effect driving observed jet angle. Assuming all sGRBs are due to compact object mergers, within a $D < 200$ Mpc aLIGO volume, we predict $0.18^{+0.19}_{-0.08}$ sGRB/GW associations all-sky per year for on-axis events at \textit{Swift} sensitivities, increasing to $1.2^{+1.9}_{-0.6}$ with the inclusion of off-axis events. We explore the consistency of our model with
GW170817/GRB~170817A in the context of structured jets. Predictions for future experiments are made.
\end{abstract}

\section{Introduction} \label{sec:intro}
Knowledge of Gamma-Ray Bursts (GRBs) has greatly benefited from the dramatic success of the \textit{Swift} satellite \citep{Gehrels_2005}, capable of rapid localization and highly-sensitive multi-wavelength observations that have enabled unprecedented ground-based followup and a high-rate of GRB redshift determinations. However, there remains relatively few short-duration GRB (sGRB) events with redshift measurements, and relatively little is understood about the properties of these populations \citep[for an overview, see;][]{Berger_2014}. 


The probable origin of sGRBs from mergers between binary compact objects such as binary neutron stars (BNS) and neutron star black hole (NSBH) binaries make them promising gravitational wave (GW) electromagnetic (EM) counterparts capable of detection by advanced interferometers. Following the success of the Laser Interferometer Gravitational-Wave Observatory (LIGO) in detecting GW150914, interest spiked in the sGRB population. The advanced aLIGO-Virgo discovery of GW170817 -- near-simultaneous with GRB~170817A detected by the \textit{Fermi} GRB Monitor (GBM) and International Gamma-Ray Astrophysics Laboratory \citep{Abbott_2017} -- provides direct evidence for BNS mergers as sGRB progenitors.

Here we seek to determine whether and how the cosmological sGRB population is consistent with
the source progenitor merger population constrained by LIGO. We employ a sample of 123 sGRBs detected by the Burst Alert Telescope (BAT) onboard \textit{Swift}, and we carefully treat the \textit{Swift}/BAT detection threshold to account for the majority of cosmological sGRBs which are undetected. We assume a universal model for the merger event rate, along with parameterized luminosity functions and beaming distributions, in order to better understand the sGRB population. The derived probabilistic sGRB world model successfully reproduces the observed \textit{Swift}/BAT number distributions in $z$ and flux, and it allows for predictions of the joint GW/sGRB detection rates for both on- and off-axis sGRBs. While the GW signal does not depend on beaming, the inclusion of distributions to describe sGRB beaming -- and the resulting number of events pointed toward the observer -- are essential to establishing a rate correspondence.

To set the stage, we carry-out an initial, approximate calculation. \textit{Swift} detects about 2 sGRBs per year within 1 Gpc, based on 7 sGRBs with measured redshift over 15.7 years and a 20\% rate of redshift determination. This corresponds to a mean, all-sky rate density for cosmological sGRBs of $20\pm 6$ Gpc$^{-3}$ yr$^{-1}$. Two aLIGO BNS events in two years of observation \citep[within 200 Mpc;][]{2021arXiv211103634T} implies a rate of 125 Gpc$^{-3}$ yr$^{-1}$, assuming constant density. Roughly then, we expect an all-sky, joint detection rate of $0.16\pm 0.05$ per year. 

Separating sGRBs from their progenitor path is difficult \citep[e.g.,][]{2022PhRvD.105h3004S},
and we will consider a rate model including both BNS and NSBH mergers.
Due to a large uncertainty in the underlying mass ratios, \cite{2021arXiv211103634T} estimates a possible rate of BNS events as large as $\approx 2000$ Gpc$^{-3}$ yr$^{-1}$, which dominates among the progenitor channels. This provides some flexibility for the source sGRB rate to be $\gtrsim 100$ times the observed rate. 
Assuming that all sGRBs are due to compact object mergers, that factor will limit the extent to which the sGRB population -- not all of which are above detection threshold -- can be beamed. We estimate a detection fraction $\lesssim 25$\% within 1 Gpc, constraining our fits to yield a mean $\theta_{\rm jet}>17$ degrees (Section \ref{sec:params}, below). We find that the observed jet angle is driven by the detection selection effect, where bias favoring bright events at increasing redshift makes the cosmological population appear quite narrowly-beamed, whereas events within the aLIGO's lower distance volume are predicted to allow a wider mean $\theta_{\rm jet}\gtrsim 30$ degrees to be seen.

\subsection{Previous Estimates} 
\label{sec:priorest}

Prior estimates of the GW/sGRB joint detection rate span a large range. Partially, the variation is
due to pre-LIGO estimates without the benefit of LIGO constraints. It is also due in large
part to differing assumptions regarding the extent of sGRB beaming.

\cite{2005A&A...435..421G} consider sGRB/BNS associations and the effect of time-delay until merger on the redshift distribution and luminosity function and find a local sGRB rate of $\sim0.8$ Gpc$^{-3}$ yr$^{-1}$. \cite{2019RAA....19..118L} find a local sGRB rate of $\sim 3-4$ Gpc$^{-3}$ yr$^{-1}$. \cite{2006A&A...453..823G} find a much higher rate of $\sim8-30$ Gpc$^{-3}$yr$^{-1}$ from redshift and luminosity distributions 
constrainted by the \textit{Swift}/\textit{HETE}-II sample. 
\cite{Guetta_2008} predict that a large fraction of detectable GW events will coincide with sGRBs. 
\cite{2006ApJ...650..281N} find a higher local rate as well of $ > 10$ Gpc$^{-3}$ yr$^{-1}$. \cite{Coward_2012} account for dominant detection biases to find an sGRB rate density of $\sim 8^{+5}_{-3} - 1100^{+700}_{-470}$ Gpc$^{-3}$ yr$^{-1}$ out to $z \approx 0.5$, assuming isotropic emission and correcting for beaming. Assuming that all BNS mergers produce an sGRB, \cite{10.1093/mnras/stt1915} estimate a simultaneous BNS-GW observation rate of $0.11-4.2$ yr$^{-1}$ with \textit{Swift} and \textit{Fermi}. Under the assumption that a notable fraction of BNS mergers result in an sGRB, \cite{Coward_2012} find a lower and upper detection rate limit for a aLIGO-Virgo search of $1-180$ yr$^{-1}$. 

Other recent studies attempt to also constrain sGRB beaming and present rates using these
assumptions.
\cite{2016A&A...594A..84G} estimate an sGRB rate within the aLIGO volume of $0.007-0.03$ yr$^{-1}$
with jet opening angles between 3 and 6 degrees. \cite{2019RAA....19..118L} estimate a similar rate of \textit{Swift} detectable sGRBs in the aLIGO horizon of $0.032$ yr$^{-1}$.
\cite{2022PhRvD.105h3004S} find a BNS merger rate of $384^{+431}_{-213}$ Gpc$^{-3}$ yr$^{-1}$, with an average $\theta_{\rm jet} \sim 15$ degrees of BNS produced sGRBs. They report that an estimated $40\%$ of BNS mergers produce jets, and measure a fraction $0.02^{+0.02}_{-0.01}$ of BNS events and $0.01 \pm 0.01$ NSBH mergers to result in observable sGRBs. \cite{2018ApJ...857..128J} use $\theta_{\rm jet} \sim 0.1$ rad and a local BNS rate density of $\sim1109^{1432}_{-657}$ Gpc$^{-3}$yr$^{-1}$, which decreases to $\sim 162^{140}_{-83}$ Gpc$^{-3}$yr$^{-1}$ when excluding a narrowly beamed sGRB. They find that off-axis sGRB events enhance the GW/sGRB rate to a $10\%$ association probability. 

\begin{figure}[hbt!]
\centering
\includegraphics[width=1\columnwidth]{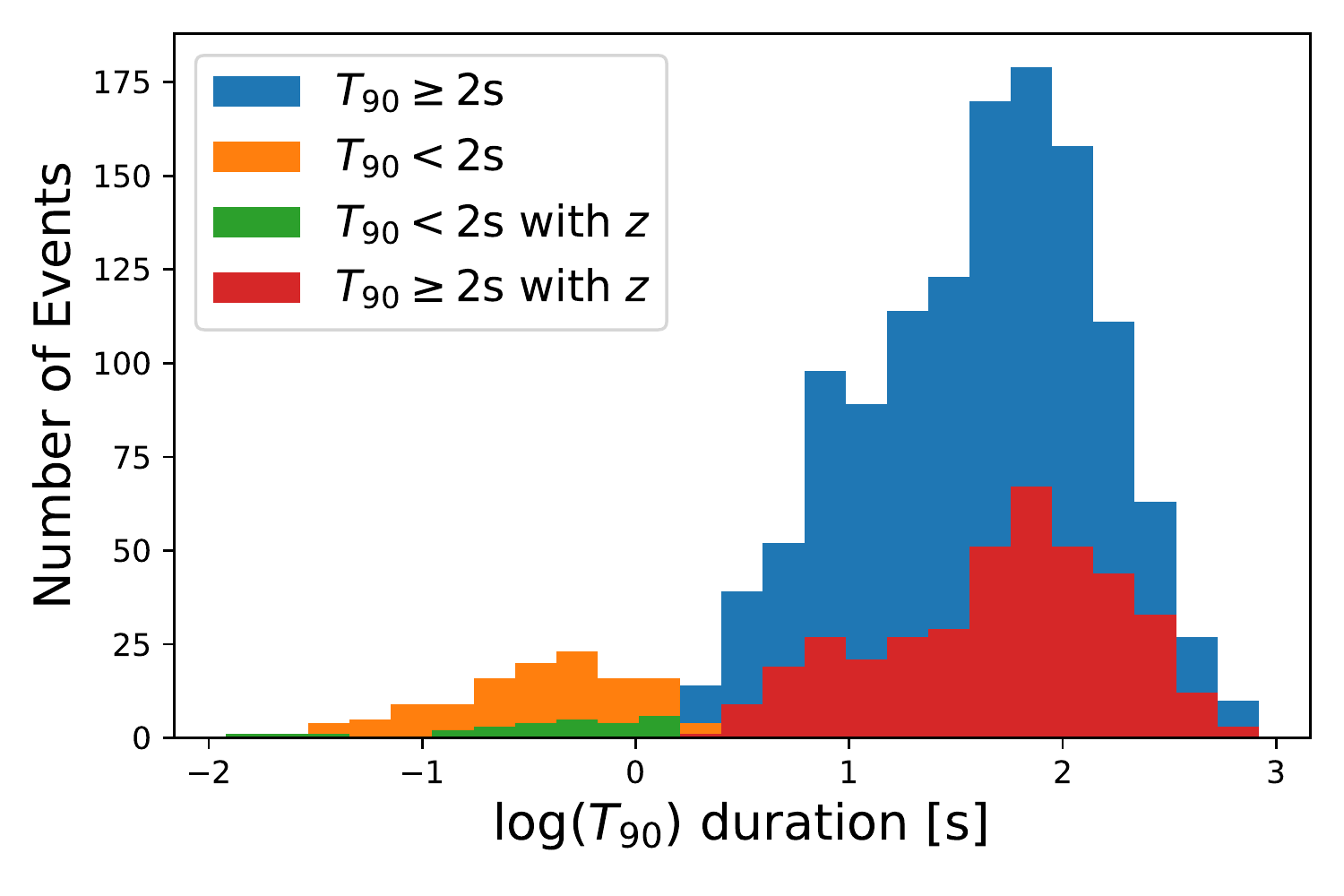}
\caption{The separation of the GRB sample into short ($T_{\rm 90}<2$s) and long-duration class, including events with and without measured $z$.}
\label{fig:logt90}
\end{figure}

\section{Data Retrieval and Sample Selection} 
\label{sec:data}

We use publicly-available GRB data from the \textit{Swift} BAT Gamma-Ray Burst Catalog\footnote{https://swift.gsfc.nasa.gov/results/batgrbcat} \citep{Lien_2016} for 1389 GRBs detected between December 17, 2004 and August 29, 2020. We retrieve burst duration intervals $T_{\rm 90}$, signal-to-noise ratios $SNR$, partial coding fractions $f_{\rm p}$, redshifts $z$ when available, and energy fluxes $F$ (15-150 keV band). A detailed description of these quantities and relevant discussion may be found in the BAT catalog or in \citet{Butler_2007}. We apply a cutoff $T_{\rm 90}<2$ s to define the short-duration burst population \citep[e.g.,][; also, Figure \ref{fig:logt90}]{1993ApJ...413L.101K}. To determine rest-frame luminosities, we assume a cosmology with h $= 0.71$, $\Omega_{\rm m} = 0.3$, and $\Omega_{\rm \lambda} = 0.7$. All error regions below correspond to the 90\% confidence intervals, unless otherwise noted.

\begin{figure*}
\centering
\includegraphics[width=\textwidth]{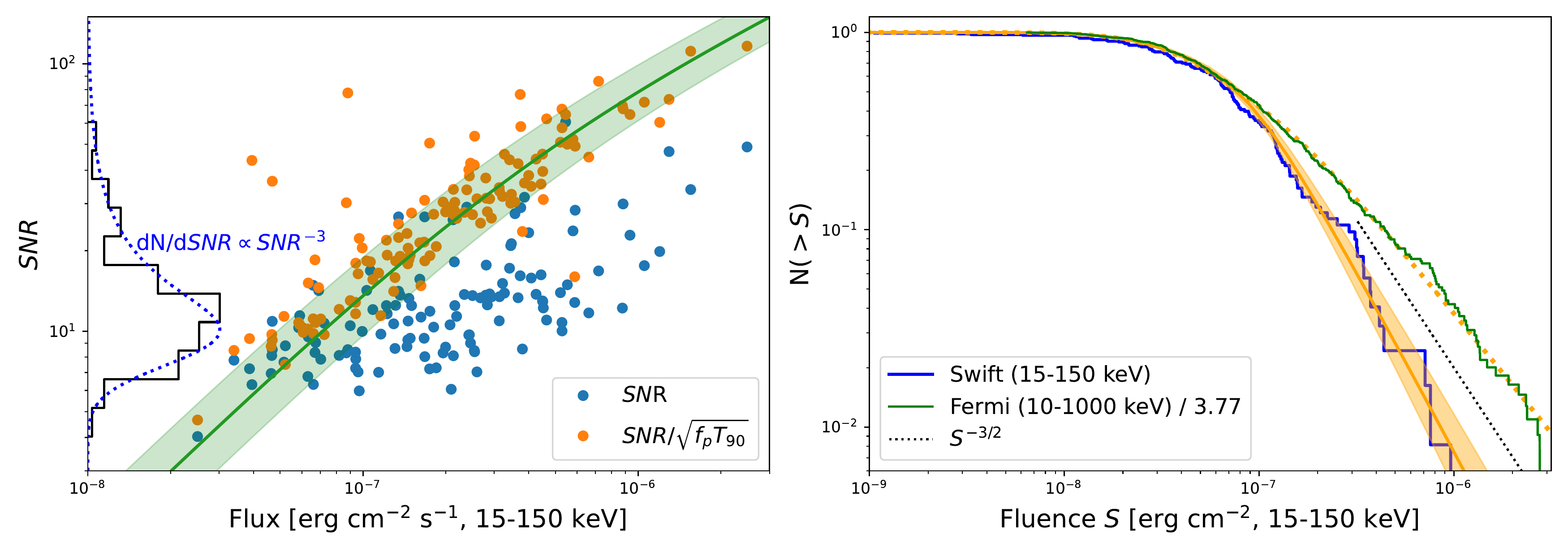}
\caption{Left: Observed $SNR$ values (blue points) -- after correction to a fixed integration time and detector area (orange points) -- correlate tightly with the observed 15-150 keV flux. The $SNR$ histogram is shown with a fitted curve that yields a detection threshold in $\log(SNR)$ of $0.9\pm 0.1$ (black; blue dashed). Right: The observed number of sGRBs over a range of Fluence $S$ for \textit{Swift} ($15-150$ keV band; blue) and \textit{Fermi} ($10-10^3$ keV band; green). The best-fit GRB world models for \textit{Swift} (orange) and for \textit{Fermi} (orange dashed) derived below are also plotted, including an $S^{0.5}$ bandpass correction factor for \textit{Swift}.}
\label{fig:detector}
\end{figure*}

A plot of the number of events versus a certain quantity can provide important lessons. The number of events with $SNR$ above a given $SNR$, is quite steep, with a slope $<-2$ (Figure \ref{fig:detector}, Left). This may suggest a steep luminosity function, but it certainly indicates -- as is well-known -- that sGRBs tend to be weakly detected. The number of events with fluence (or flux) above a certain level, the so-called logN-logS relation, can be used to to infer the nature of the source population. A logN-logS slope of $-3/2$ for a relatively nearby population (such as sGRBs) is only weekly-dependent upon the luminosity function and is indicative of a non-evolving source population \citep[e.g.,][]{1966MNRAS.133..421L}. Such a slope appears to be present at the bright end of the logN-logS relation (Figure \ref{fig:detector}, Right). The faint-end slope for sGRBs may also be near $-3/2$ after accounting for threshold effects 
\citep[e.g.,][]{1996AIPC..366..170P}. However, to disentangle the physical effects that drive the number counts -- to separately measure the luminosity function and the number density versus redshift -- redshift measurements and a careful treatment of the detection limit are required. Parametric modelling is beneficial given the typically low $SNR$ values.

In order to use the measured data to constrain intrinsic GRB properties, we need a prescription for mapping rest frame luminosities to observer frame fluxes and for deciding whether the observed fluxes are above the BAT detection threshold. We find that $SNR$ for \textit{Swift} can be predicted from the flux $F$ with 0.1 dex scatter (i.e., $1\sigma$ spread; see, Figure \ref{fig:detector}, Left), as:
\begin{equation}
    SNR =  (F/F_{\circ}) \sqrt{ T_{\rm 90} f_p } / \sqrt{ 2950 + F/F_{\circ} }.
    \label{eq:snr}
\end{equation}
Here, the zero point $F_{\circ} = 1.2 \times 10^{-10}$ erg/cm$^2$/s, and 2950 represents a typical background count rate. We note that the observed $SNR$ distribution is very steep ($dN/dSNR \propto SNR^{-3}$) -- indicating relatively little dynamic range in the sample -- and is well-fit by a powerlaw truncated to zero at threshold and convolved with a Gaussian. This yields a stochastic $SNR$ threshold of $10^{0.9\pm 0.1}$.

\begin{figure*}
\centering
\includegraphics[width=\textwidth]{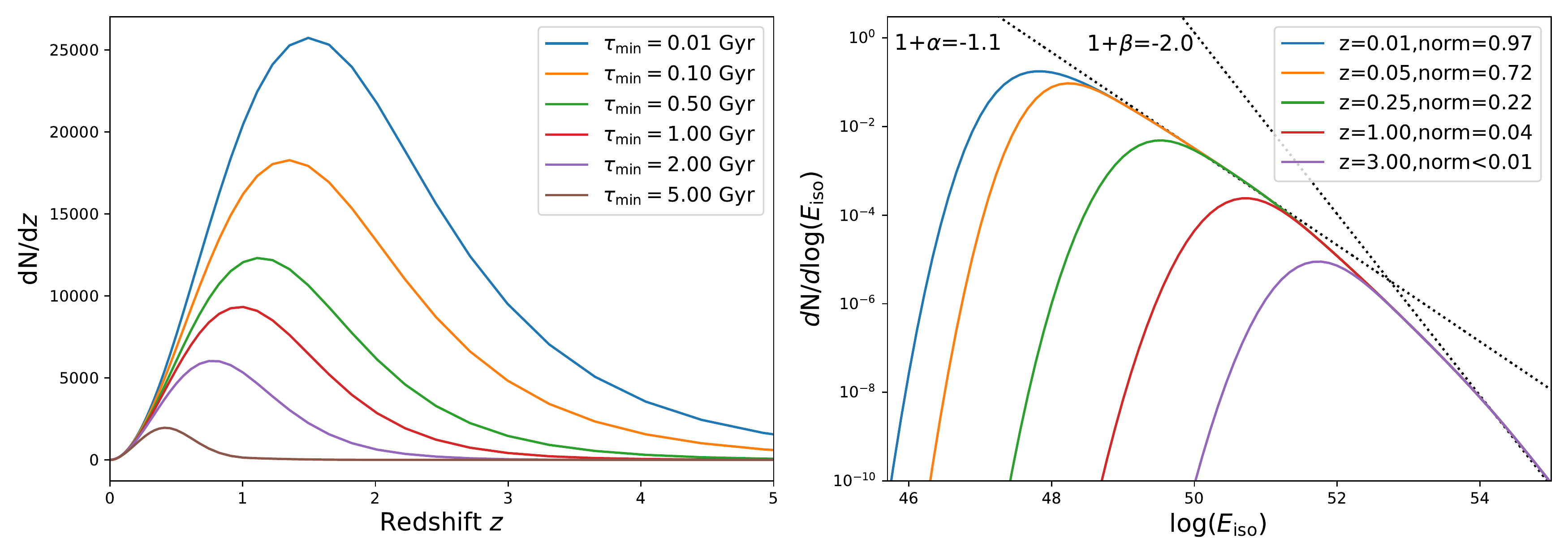}
\caption{Left: Binary merger rates as a function of $z$ from \citet{Safarzadeh_2019} for different minimum time delays $\tau_{\rm min}$ between formation -- which is assumed to track the SFR and Universal volume -- and merger. Right: The $E_{\rm iso}$ luminosity function -- found by integrating Equation \ref{eq:E_alphabeta} for the $L_{\gamma}$ luminosity function over $f_b$ -- includes the detector threshold which increases with increasing $z$. The asymptotic power law behavior is shown with dotted lines. The integral of the model (see ``norm'' in legend) provides the fraction of sGRBs detected.}
\label{fig:model}
\end{figure*}

We assume flux $F$ and fluence $S$ in the \textit{Swift}/BAT bandpass are related as $S=F T_{\rm 90}$. Due to the low $SNR$ values and the soft \textit{Swift} bandpass, spectral parameters are typically poorly-constrained and the conversion to bolometric $S$ is challenging. To simplify matters, we assume $S_{\rm bol} = x S$, where $x$ is a bandpass conversion factor applied to all bursts. We expect a large uncertainty of about a factor of 2 (0.3 dex) for $x$. We assume a mean correction $x=3.77$ which is suitable for a typical, soft \textit{Swift} sGRB spectrum ($E_{\rm peak}\approx 250$ keV). A change to this value would alter the energetics scale (e.g., shift locations on some plots below). 

This simple approach to a bolometric correction appears to break down for bright sGRBs (see, Figure \ref{fig:detector}, Right). In comparison with sGRB fluences from \textit{Fermi}/GBM
\citep{Gruber_2014,vonKienlin_2014,Bhat_2016,vonKienlin_2020},
we find that a larger correction is required for bright bursts. For bursts with $S>10^{-7}$ erg/cm$^2$ (15-150 keV), we assume that the bolometric correction factor increases as $S^{0.5}$. This observer-frame effect likely arises due to the increasing hardness of bright bursts and the resulting lack of flux in the soft 15-150 keV band as compared to the harder, and approximately-bolometric 10-1000 keV \textit{Fermi}/GBM band. Although the effect is confirmed to correlate with hardness in the \textit{Swift} 50-150 keV to 15-50 keV bands for 12 bursts measured in common between \textit{Swift} and \textit{Fermi}, we model it indirectly (on $S$ alone) because a correction using the \textit{Swift} hardness would introduce large errors and require spectral modelling not possible (with sufficient accuracy) given only the soft \textit{Swift} bandpass. While a bandpass correction based on $S$ only would not be an acceptable approach for individual burst modelling, it does capture the ensemble bandpass effects well. From Figure \ref{fig:detector} (Right), we see that both \textit{Swift} and \textit{Fermi} have similar sGRB sensitivities \citep[see also,][]{Band2003}, and both have logN-logS distributions scaling roughly as $S^{-3/2}$ at the bright end.

Finally, we note that there are 3 events with anomalously high luminosities ($>10^{52}$ erg/s) which we exclude from the analysis, but return to in Section \ref{sec:model_fitting} below. GRBs~090426 and 111117A have spectroscopic $z>2$ and may lie on the short-duration end of the long-duration GRB population \citep[see,][]{refId0,10.1111/j.1365-2966.2009.15733.x,selsing}. GRB~120804A has an uncertain photometric $z=1.3^{+0.3}_{-0.2}$ (1 sigma) based on a study of the host galaxy \citep{Berger_2013}. Our final sample includes 123 sGRBs between and including GRB~050502 and GRB~20022A. Of these, 27 (22\%) have measured redshifts in the catalog. 

\section{sGRB World Model Summary} 
\label{sec:models}

Following \citet{Butler_2010}, to model the observed rates (number versus luminosity, number versus redshift $z$, number versus beaming fraction, etc.), we define functional forms for the probability distributions and, for a given set of the parameters, use the intrinsic model to generate sGRBs reaching the detector. Not all of these will be detected, and evaluation of the extent to which the predicted rates from a given set of parameters match the observed rates requires the detector model above (e.g., Figure \ref{fig:detector}).

We assume a sGRB rate density model $\dot \rho(z)$ due to compact object mergers from \cite{Safarzadeh_2019}, which is a convolution of the star formation rate (SFR) and a merger delay time $\tau_{\rm d}$ distribution. To convert from rate density to number $N$ per redshift interval: 
\begin{equation}\label{eq:dndz}
\vspace{-0.05in}
P_z(z) = {dN \over dz} = \dot \rho(z) {dV \over dz} { 1 \over 1+z}.
\end{equation}
Here, $V$ is the Universal volume at $z$, and the factor of $1+z$ converts rates at $z$ to now. We assume a $\tau_{\rm d}$ distribution $\propto \tau_d^{-1}$, where the maximum delay time is 10 Gyr and the minimum delay time ($\tau_{\rm min}$) is a free-parameter. In principle, the $-1$ slope can also be varied; however this provides no additional model flexibility over the $z$ range under study here. Resulting $dN/dz$ distributions are shown in Figure \ref{fig:model} (Left). The rate density tracks the SFR for $\tau_{\rm min}\lesssim 100$ Myr and peaks at lower $z$ for larger $\tau_{\rm min}$.

The sGRB luminosity function is generally modelled as a single or broken power law, similar to studies of long-duration GRBs \citep[see, e.g.,][]{2006A&A...453..823G, 2008MNRAS.388L...6S,Virgili_2011,2014MNRAS.442.2342D,2016A&A...594A..84G}.
To get the number of GRBs over a range of $\gamma$-ray Luminosity $L_{\gamma}$, we assume $P_L=dN/dL_{\gamma}$, and
\begin{equation}\label{eq:E_alphabeta}
\begin{split}
E_{\circ} { dN \over dL_{\gamma}} = A (L_{\gamma}/E_{\circ})^{\alpha}, \; \; \; \; E_{\rm min}<L_{\gamma}<E_{\circ}
\\ = A (L_{\gamma}/E_{\circ})^{\beta}.  \; \; \; \; \; \; \; \; \; \; \; \; \; \; \; \; \; L_{\gamma} \geq E_{\circ} 
 \end{split}
\end{equation}

The luminosity function is defined above a minimum $E_{\rm min}$, allowing for it to be normalized (fixing $A$) assuming $\alpha\ge\beta$ and $\beta<-1$. 

Choices for physical quantities to associate with $L_{\gamma}$ include the total $\gamma$-ray energy reservoir, otherwise known as the beaming-corrected energy release $E_{\gamma}$, or the energy per solid angle as represented by the isotropic-equivalent energy release $E_{\rm iso}$. Here, $E_{\gamma} = f_b E_{\rm iso}$, for beaming fraction $f_b = 1 - \cos(\theta_{\rm jet})$. Also, we are assuming the simple top-hat GRB jet model \citep[e.g.,][]{Rhoads1999ApJ,2001ApJ...562L..55F}, where the GRB is detected only if it launches a jet pointed in the direction of the observer. The rate of such events is proportional to the fraction of sky illuminated (i.e., $f_b$), establishing a direct relationship between outflow collimation and rates \citep[see, also,][]{Berger_2014}. 

To allow some flexibility between these possibilities, we write $L_{\gamma} = E_{\rm iso} f_b^c$. The index $c$ represents a potential correlation between $f_b$ and $E_{\gamma}$. We allow $c$ to vary between 0 and 1, allowing the luminosity argument to vary between $E_{\rm iso}$ ($c=0$) and $E_{\gamma}$ ($c=1$). Viewed differently, the luminosity function is allowed to regulate GRBs in terms of total energetics ($c=1$) or local properties ($c=0$). We can motivate allowing this flexibility by noting that, during a GRB, the outside of the jet is not in causal contact with the inside of the jet; the processes that govern the observed GRB brightness distribution could depend weakly on the global energetics. 

Since detection depends on $E_{\rm iso}$, the observed $f_b$ distribution will change with $z$ for $c>0$. With $c=0$, the luminosity function depends on $E_{\rm iso}$, and the observed $f_b$ will take on an average value for all $z$. We assume a powerlaw $P_{f_b} = dN/df_b$ distribution \citep[as in, e.g.,][]{2001ApJ...562L..55F}, with
\begin{equation}
    dN/df_b = B f_b^n
    \label{eq:fb_dist}
\end{equation}
The normalization $B$ is set by integrating between a minimum $f_{b,{\rm min}}=10^{-4}$ (corresponding to $\theta_{\rm jet,min}=0.8$ degrees) and a maximum $f_b=1$. We apply a weak constraint on the beaming distribution: 
\begin{equation}\label{eq:n_constraint}
    n < -1.5 - c \; (1+\alpha).
\end{equation}
This prevents a flat, observed $\theta_{\rm jet}$ distribution and corresponds to the statement ``short-duration GRBs are beamed.'' 

We assume a log-normal distribution of $\log(T_{\rm 90}) = -0.5 \pm 0.4$ and an exponential distribution for $\log(f_p)$ with mean -0.23 as in \cite{Butler_2010}. 

Combining the above distributions, the predicted, observed event rate density $r$ is:
\begin{equation}\label{eq:r}
\begin{split}
r = r_{\circ} f_b P_{f_b}(f_b|n) P_z(z|\tau_{\rm min}) P_L(L_{\gamma}|\alpha,\beta,c,E_{\rm min},E_{\circ}) \\ 
  \times \;  H(L_{\gamma}-L_{\rm th}) P_{th}(L_{\rm t}|L_{\gamma,{\rm min}}).
\end{split}
\end{equation}
Here, $L_{\gamma} = 4\pi D_L^2/(1+z) x S f_b^c$, and the minimum detectable $L_{\gamma,{\rm min}} = 4\pi D_L^2/(1+z) x S_{\rm min}(T_{\rm 90},f_p) f_b^c$, and $H$ is the Heaviside step function. The threshold $L_{\rm th}$ distribution $P_{th}(L_{\rm th})$ is assumed to be log-normal with a mean of $L_{\gamma,{\rm min}}$ and a standard deviation of 0.35 dex. 
This accounts for the uncertain conversions from limiting $SNR$ to limiting \textit{Swift} 15-350 keV $S$ (0.14--0.18 dex, depending on whether we fix $T_{\rm 90}$ and $f_b$ or average over the observed values) and then from \textit{Swift} $S$ to bolometric $S$ (0.3 dex). The contribution to the scatter from averaging over $T_{\rm 90}$ and $f_p$ is negligible, and we can effectively marginalize over those distributions when calculating the distribution normalization by assuming their mean values.

\subsection{Model Normalization}
\label{sec:norm}

We assume all sGRBs are due to compact object mergers. The normalization $r_{\circ}$ in Equation \ref{eq:r} includes 3 factors, 1 fixed and 2 variable which become model parameters requiring priors. First is the fixed $\Omega/(4\pi)$ fraction of sky covered by \textit{Swift}, for which we assume $\Omega$ = 1.3 str, multiplied by a \textit{Swift} observation period of 15.7 years. Second is the rate of merger events in the $D<200$ Mpc aLIGO volume. This normalization factor is represented by a gamma function prior corresponding to 2 BNS events 
\citep[GW170817 and GW190425,][]{PhysRevLett.119.161101,2020ApJ...892L...3A} detected in 2 years (O1 through O3). Third is a factor that scales the $D<200$ Mpc yearly rate out to a $D=1$ Gpc yearly rate appropriate for sGRBs. The third normalization factor is highly-uncertain based on LIGO team estimates \citep{2021arXiv211103634T} and is assumed to be distributed uniformly between 1 and 10. We choose a distance of 200 Mpc as a representative LIGO sensitivity volume for BNS
events; LIGO is expected to detect BNS events out to 190 Mpc in the O4 observing run 
(2023--2024) and out to 325 Mpc in 
O5 (2026--2028)\footnote{https://observing.docs.ligo.org/plan}.

\begin{figure*}[t!]
\centering
\includegraphics[width=1.01\textwidth]{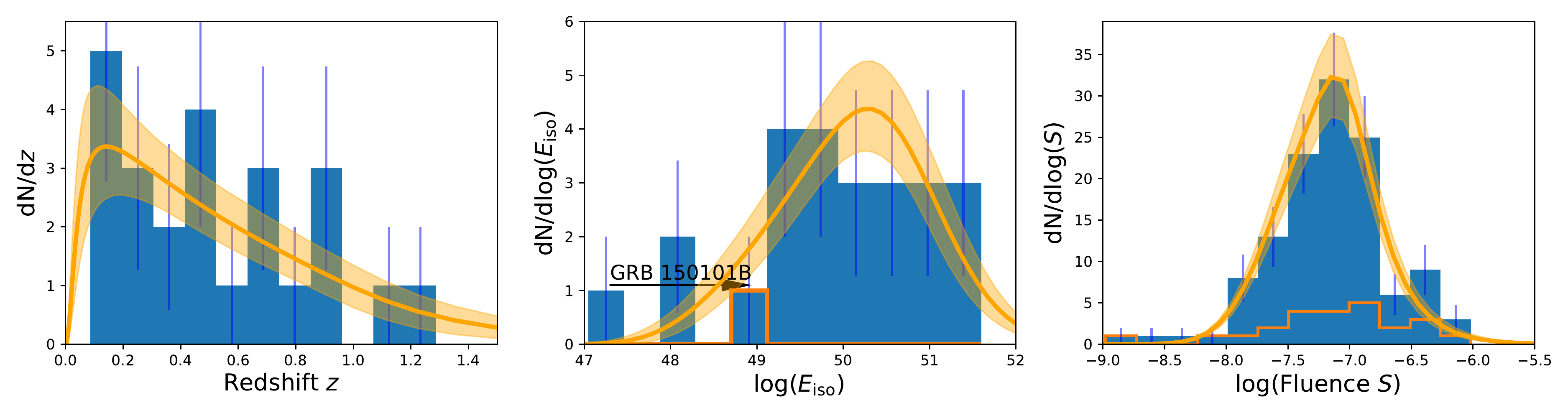}
\caption{The observed (blue) and predicted (orange) rates of sGRBs with respect to $z$ (Left), $E_{\rm iso}$ (Middle), and $S$ (Right). The right panel includes all sGRBs in the blue histogram and also displays the sGRBs with measured $z$ as a superimposed, unshaded orange histogram. GRB~150101B is discussed in Section \ref{sec:obsdist}.}
\label{fig:obshisto}
\end{figure*}

\subsection{The Distribution of $E_{\rm iso}$, given $z$}\label{sec:lum}

To fit the data, we require $P_E$: the distribution of $E_{\rm iso}$, given $z$. We construct $P_E$ by multiplying the luminosity function by the beaming distribution, and the detection distributions by $f_b$, evaluating at $L_{\gamma} = f_b^c E_{\rm iso}$, and integrating over $f_b$. The integration over $f_b$ results in a smoothly-broken powerlaw with low-energy slope $\alpha$ and high-energy slope $\beta$ (see, Figure \ref{fig:model}; Right). The transition between slopes occurs over an extended range in $E_{\rm iso}$ from $E_{\circ}$ until $E_{\circ} / f^c_{b,{\rm min}}$. To be explicit, starting from Equation \ref{eq:r}:
\begin{equation}\label{eq:r2}
r = r_{\circ} P_z(z|\tau_{\rm min}) P_E(E_{\rm iso}|z,n,\alpha,\beta,c,E_{\rm min},E_{\circ}),
\end{equation} with
\begin{equation}\label{eq:P_E}
P_E = \int
f_b P_{f_b} P_L(E_{\rm iso} f_b^c) H(E_{\rm iso}f_b^c-L_{\rm thresh}) P_t df_b.
\end{equation}

We first calculate $P_E$ analytically at $z=0$ (i.e., all events detected), and ensure that it is normalized. The detection distributions $H P_t$ are then applied as a $z$-dependent cutoff at $L_{\rm thresh}=E_{\rm iso}f_b^c$, followed by a Gaussian convolution to get the observed distribution in $E_{\rm iso}$ at a given $z$. 

From $P_E$ and $P_z$, it is then possible to calculate the joint distribution of $S$ and $z$ and the marginal distributions for $E_{\rm iso}$, $S$, or $z$. The integrations over $E_{\rm iso}$, $S$, and $z$ are carried out numerically over fine grids, with the $z$ grid extending from $z=7.1 \times 10^{-4}$ (i.e., $D=3$ Mpc) to $z=10$.
Figure \ref{fig:model} (Right) displays the $E_{\rm iso}$ luminosity function at different $z$ values.

\subsection{Model Fitting}
\label{sec:model_fitting}

In fitting the above models by matching the model and observed rates, we treat the measured values of $z$, $T_{\rm 90}$, $f_p$, and $S$ (and $E_{\rm iso}$ fixed by $z$ and $S$) as point estimates only (i.e., having no error). The uncertainty in these quantities is placed in the models, which can be considered smoothed to avoid over-fitting.

The parameters $\vec{\theta}$ are $n$, $c$, $E_{\rm min}$, $\alpha$, $\beta$, $E_{\circ}$, and $\tau_{\rm min}$. The observables $\vec{D}$ are $z$, $S$, $T_{\rm 90}$, and $f_p$. The rate evaluated for a given event $i$ is written $r_i$. In the case of GRBs with no measured $z$, we integrate $r_i$ over $z$.
To find the parameters that best fit the model, we maximize the Poisson likelihood:
\begin{equation} \label{eq:L}
\mathcal{L}(\vec{D}|\vec{\theta}) = \exp\left[- \int r d\vec{D}\right] \prod_{i=1}^N r_i(\vec{\theta})
\end{equation}

The integral in the exponential in Equation \ref{eq:L}, with $d\vec{D} = dz\,dS\,dT_{\rm 90}\,df_p$, yields the total number of predicted events.
Prior to crawling the parameter space $\vec{\theta}$ via Markov Chain Monte Carlo (MCMC) maximization of $\log(\mathcal{L})$, we first marginalize analytically over $r_{\circ}$ (see, Equation \ref{eq:r2}).

The MCMC analysis is carried out with PyMC3 \citep{pymc}, using multiple chains at a variety of starting
parameter values to establish uniqueness of the derived solution. We assume uniform prior distributions on all parameters except for $E_{\rm min}$ and $E_{\circ}$. Also included are the luminosity function slope constraints (Section \ref{sec:models}) and constraints on the $f_b$ distribution (Equation \ref{eq:n_constraint}). For $E_{\rm min}$ and $E_{\circ}$ we assume log-normal priors, $\log(E_{\rm min}) = 47.8 \pm 0.5$ and $\log(E_{\circ})=50.5\pm 0.3$. The $\log(E_{\rm min})$ prior is weak and non-constraining. The tighter $E_{\circ}$ prior is designed to test for the possibility of a break in the luminosity function.

\begin{figure*}
\centering
\includegraphics[width=0.9\textwidth]{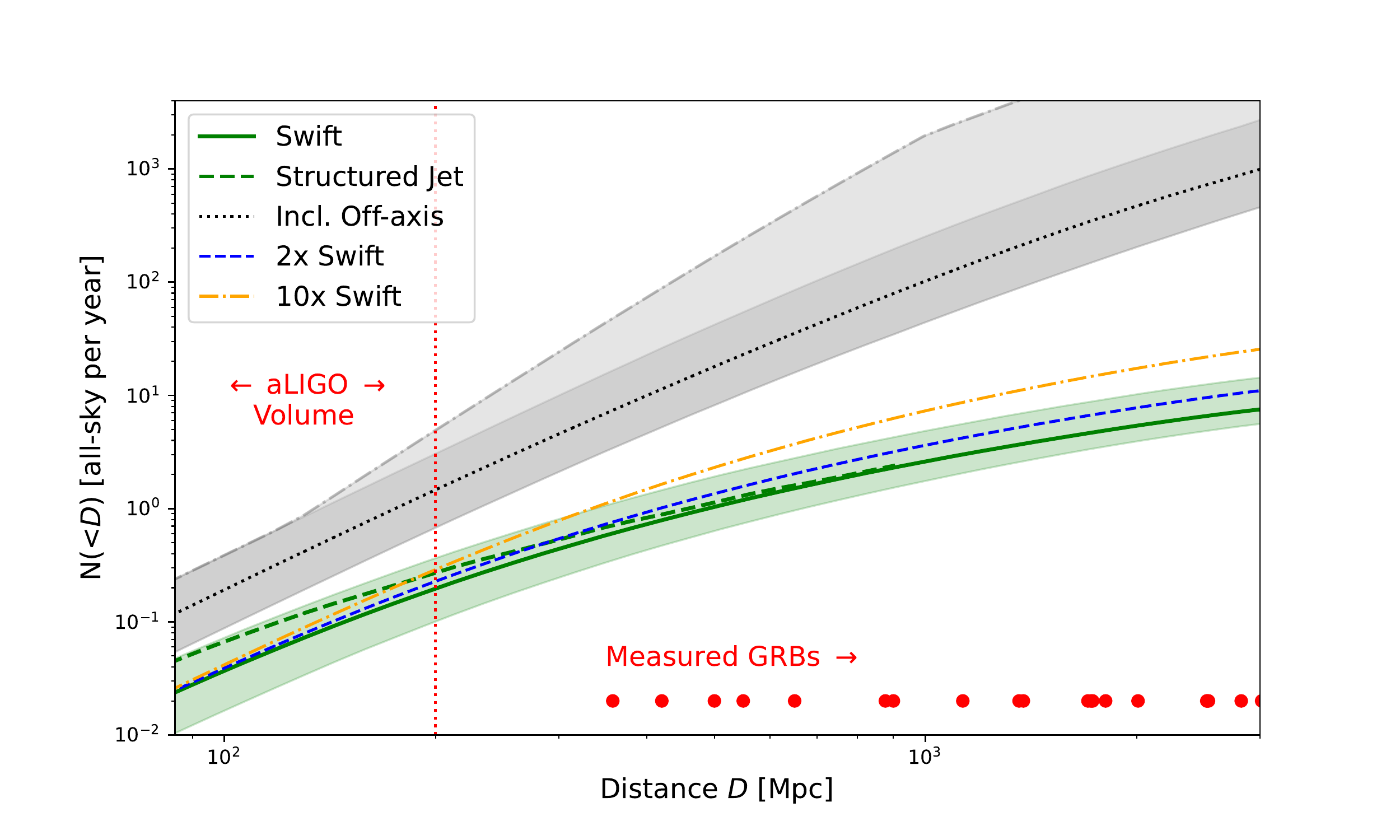}
\caption{The all-sky predicted number of sGRBs per year as a function of distance, detectable by \textit{Swift}. On-axis events are displayed in green, and on- and off-axis bursts are displayed in black. Increased detector area (blue and orange curves) strongly-affect the predicted rate at $D\gtrsim 1$ Gpc. Structured jets (dashed green curve) are
discussed in Section \ref{sec:conclusion}.}
\label{fig:ligo}
\end{figure*}

\begin{table}[hbt!]
\centering
\caption{MCMC Parameters 90\% Confidence Intervals}
\begin{tabular}{ll}
\hline
Parameter                         & Value (all GRBs)     \\ \hline
$f_b$ index $n$                       & $-0.7 \pm 0.3$ \\
correlation index $c$                        & $>0.6$  \\
$\log(E_{\rm min})$                   & $47.3 \pm 0.4$ \\
$\alpha$             & $-2.1\pm 0.2$ \\ 
$\beta$              & $-2.5^{+0.3}_{-2.1}$ \\
$\log(E_{\circ})$                   & $50.5^{+0.6}_{-0.3}$  \\
$\tau_{min}$            & $<700$ Myr \\
\hline
\end{tabular}
\label{tab:table-param}
\end{table}

The posterior parameter distributions are estimated by drawing $10^4$ parameter $\vec{\theta}$ samples from the MCMC, after thinning by a factor of 4 to limit draw-to-draw correlations. These are summarized in Table \ref{tab:table-param} and discussed more in Section \ref{sec:params} below. We generate posterior distributions for the observables $\vec{D}$ by averaging their distributions from Section \ref{sec:lum} over the sampled parameters.

\section{Discussion}
\label{sec:discussion}

\subsection{Predicted Observable Distributions}\label{sec:obsdist}

Predicted distributions and uncertainties for $z$, $E_{\rm iso}$, and $S$ are overplotted on the observed data in Figure \ref{fig:obshisto}. The models fit the observed data well. The best-fit fluence model (Figure \ref{fig:obshisto}; Right), for example, has a reduced $\chi^2_{\nu} = 1.1$ for $\nu=4$ degrees of freedom, using the 
\citet{Gehrels1986ApJ} approximate for the standard deviation of Poisson counts.

While the observed redshift distribution peaks at $z\approx 0.2$, this is largely
due to the decline in detection efficiency at high-$z$ (Figure \ref{fig:model}; Right).
The input redshift distribution -- Equation \ref{eq:dndz} averaged over draws
of the $\tau_{\rm min}$ parameter -- peaks at $z=1.35$, as compared to $z=1.45$ for the SFR.
We estimate a modest, 73\% confidence that the rate model is delayed with respect to the SFR
by counting the fraction of draws with a peak below $z=1.4$.

The predicted distributions for $E_{\rm iso}$ and fluence $S$ are also closely
consistent with the data. However, we note that in the middle panel of Figure \ref{fig:obshisto}, there is a possible outlier. GRB~150101B was an un-triggered event also detected by \textit{Fermi}/GBM with a significantly higher reported flux (leading to $E_{\rm iso}$ = $1.1 \times 10^{49}$ erg instead of $1.6 \times 10^{47}$ erg \citep{Burns_2018}. 
GRB~150101B may have also been an off-axis event \citep{Troja2018}.

In the fitting above, we have assumed that events with $z$ should be fit alongside events without redshift. This can be justified by the similarity of the observed $S$ distributions in Figure \ref{fig:obshisto} (Right). From a two-sample Kolmogorov-Smirnov test, we find that the distributions cannot be significantly differentiated ($\tau=0.168$; p-value 0.58).

\begin{figure*}
\includegraphics[width=\textwidth]{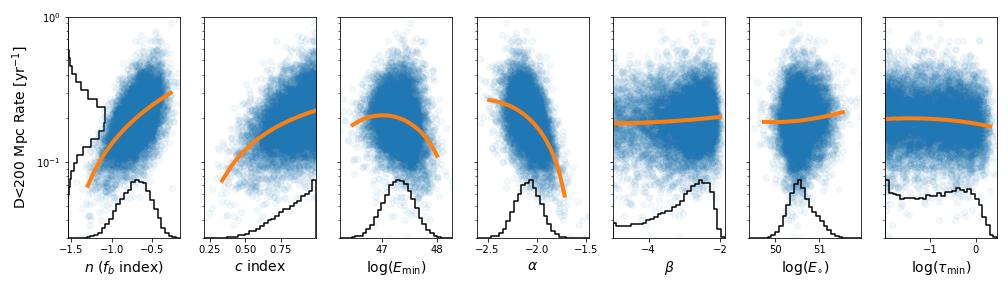}
\caption{The number of on-axis sGRBs expected within the aLIGO volume, displayed versus the different parameters. Quadratic curves are plotted through the parameter samples to highlight possible trends. Overall distributions are displayed as histograms on both axes (see also, Table \ref{tab:table-param}).}
\label{fig:ligoparams}
\end{figure*}

The evidence for a break in the luminosity function is formally strong (99\% confidence), given the confidence intervals for $\alpha$ and $\beta$. However, the measurement of $E_{\circ}$ without the restrictive prior is highly uncertain given the smooth shape of the luminosity function that results from the uncertainty in our $E_{\rm iso}$ measurements and the additional smoothing caused by integration over $f_b$. Moreover, it is important to note that the possible measurement of a break depends on the sample selection. If we include the 3 high-luminosity events in our fitting (GRBs 090426, 111117A, and 120804A; Section \ref{sec:data}), the slopes become entirely consistent ($\alpha = -2.1^{+0.2}_{-0.1}$, $\beta = -2.2 \pm 0.1$) with a single power-law luminosity function. Also, including these events lowers the upper limit on $\tau_{\rm min}$ (30 versus 700 Myr). The other model parameters in Table \ref{tab:table-param} are not significantly altered.

\subsection{GRB Rate Predictions}
\label{sec:LIGO}

The integral of Equation \ref{eq:P_E} over $E_{\rm iso}$ at a given $z$, averaged over MCMC parameters (Table \ref{tab:table-param}), provides the total, predicted event rate for on-axis sGRBs as a function of redshift. If we omit the first $f_b$ factor from the integral, we can determine the predicted event rate for all sGRBs on- or off-axis as a function of redshift. These curves and their uncertainties given the spread in model parameters are displayed in Figure \ref{fig:ligo}. The curves are generated assuming the \textit{Swift} sensitivity but all-sky coverage. 

At large distances ($D\gtrsim 1$ Gpc), the satellite sensitivity becomes important as many events are lost below threshold. Curves for hypothetical satellites with sensitivities two times and ten times greater than \textit{Swift} are displayed. For distances below 1 Gpc and above 200 Mpc, the dominant rate uncertainty is due to the uncertainty in the total number of BNS mergers as constrained by LIGO \citep{2021arXiv211103634T}. This uncertainty is displayed using a dot-dashed curve at the top of Figure \ref{fig:ligo}. 

Below 200 Mpc, essentially all sGRBs are detected by \textit{Swift} according to our modelling.
The number of events detectable all-sky per year at \textit{Swift} sensitivities within the aLIGO volume is $0.18^{+0.19}_{-0.08}$ for on-axis events -- in agreement with our rough calculation in Section \ref{sec:intro} -- and increases to $1.2^{+1.9}_{-0.6}$ with the inclusion of off-axis events.

\subsection{Rate Dependence on Beaming, Other Parameters}
\label{sec:params}

In Figure \ref{fig:ligoparams}, we display the predicted rate determined for each parameter draw, with a separate sub-panel dedicated to each parameter. A quadratic curve is fit in each sub-panel to highlight trends. Significant trends are observed for 3 parameters: $n$, $c$, and $\alpha$. 

When the low-energy powerlaw index for the luminosity function $\alpha$ is steep, there is naturally an increased rate of faint, nearby events. This effect can be observed weakly in the $E_{\rm min}$ parameter as well. However, the impact of that parameter is limited by the fact that $E_{\rm min}$ values below the observed range of measured $E_{\rm iso}$ values would lead to an over-large model normalization and be penalized by the exponential term in Equation \ref{eq:L}.

Several of the parameters are correlated with each other as a result of the flux
limit and the beaming constraint Equation \ref{eq:n_constraint}. The rate-density parameter, however,
is uncorrelated with the other parameters, except for a weak correlation with
$\alpha$ for long delay times ($\tau_{\rm min}>500$ Myr).

We note that our derived luminosity function slopes $\alpha \approx -2$ are considerably steeper than $\alpha\approx -0.5$ found in \cite{2016A&A...594A..84G}. While we do not claim strong evidence for a break in the luminosity function (Section \ref{sec:obsdist}), and therefore a clean separation between the slopes $\alpha$ and $\beta$, we do find that the overall luminosity function is steep. 

\begin{figure*}
\centering
\includegraphics[width=\textwidth]{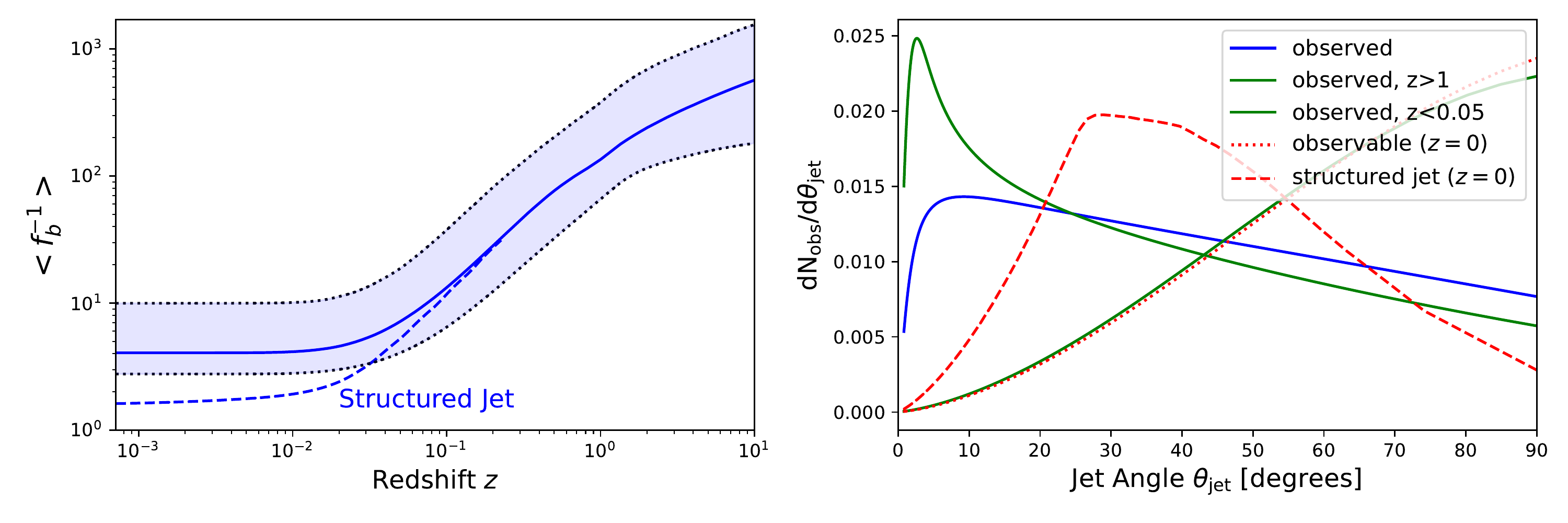}
\caption{Left: Mean inverse beaming fraction $\langle f_b^{-1} \rangle$ -- proportional to the number of missed events -- and confidence intervals (dotted lines, blue shaded regions) of sGRBs as a function of $z$. Right: The mean distribution of sGRB jet angles dN$_{\rm obs}$/d$\theta_{\rm jet}$. We show the predicted observed (blue, $z > 1$ green) distributions as well as those expected for an ideal instrument (i.e., $z=0$; red dotted). Structured jets (dashed curves) are discussed in Section \ref{sec:conclusion}.}
\label{fig:beaming}
\end{figure*}

The steeper luminosity function likely arises from our approach to the modelling of beaming. 
A large, negative value for $n$ corresponds to a highly-beamed population where the bulk of sGRBs are pointed away from the observer. It is not possible to generate the observed sGRB population from compact object mergers if $n$ is too negative. However, with $f_b$ present in the luminosity function, an interesting effect arises where a steep $\alpha<-1$ permits a larger $n$ (i.e., more GRBs pointed at the observer; see, Equation \ref{eq:n_constraint}).

The effect of a correlation index $c$ near unity is to make the luminosity function depend on $E_{\gamma}$. The data prefer a luminosity function depending on $E_{\gamma}$ rather than $E_{\rm iso}$ ($c>0.5$ at 98\% confidence). In this case, there arises a significant detection bias in favor of narrowly-beamed events with increasing redshift (see, Figure \ref{fig:beaming}). 
The observed cosmological sGRB population (i.e., at high-$z$) appears narrowly beamed (satisfying the constraint Equation \ref{eq:n_constraint}), while the low-$z$ population tends to be more broadly beamed, leading to a higher predicted rate. 

Direct constraints on sGRB jet opening angles from afterglow jet-break measurements are rare.
\cite{2014PhDT_fong} present a complete tabulation of sGRB beaming angles measurements and limits available prior to 2014. From 4 measurements of jets with $\theta_{\rm jet} \approx 5$ degrees and 9 upper limits (as large as 25 degrees), \cite{Berger_2014} estimate a mean $\theta_{\rm jet} > 10$ degrees. Our predictions are consistent with these constraints (see, Figure \ref{fig:beaming}; Right), allowing for large beaming angles but implying that the cosmological population should appear significantly more tightly beamed than the general population. We find a most-likely $\theta_{\rm jet} = 2.6$ degrees for $z>1$ as compared to $\theta_{\rm jet} = 9$ degrees for all $z$.

An additional important model parameter not displayed in Figure \ref{fig:ligoparams} is the
normalization. To reproduce all cosmological sGRBs we must exploit
the large (factor of 10) uncertainty in the BNS rate within 1 Gpc (Sections \ref{sec:intro}, 
\ref{sec:norm}). We find that this factor must be $>5$ (90\% confidence), corresponding
to a BNS rate $>10^3$ Gpc$^{-3}$.

\section{Conclusions}\label{sec:conclusion}

We have demonstrated that a sample of 123 sGRBs from \textit{Swift} -- including 27 events with measured redshifts -- can be generated from the source aLIGO merger population, in turn deriving 
probabilistic estimates for the sGRB rate density, the luminosity function, and the
beaming distribution. The rate-density appears to allow for a range of merger delay
times ($\tau_{\rm d,min}<700$ Myr) and suggests contributions from BNS and NSBH progenitor channels \citep[e.g.,][]{Safarzadeh_2019,2022PhRvD.105h3004S}, although we have not attempted to disentangle these contributions. The luminosity function is a relatively steep ($\alpha\approx -2$), 
possibly-broken powerlaw operating on $E_{\rm iso}$ and the beaming fraction $f_b$ and 
leading to a correlation between the two. As a result, we predict relatively wide sGRB jet angles, particularly for nearby events.

We predict $0.18^{+0.19}_{-0.08}$ GW/sGRB associations per year, all-sky, for on-axis events at \textit{Swift} sensitivities within the $D< 200$ Mpc aLIGO volume. This increases to $1.2^{+1.9}_{-0.6}$ yr$^{-1}$ with the inclusion of off-axis events. These very nearby events have a broad, jet-opening angle
distribution, with mean $\theta_{\rm jet}\gtrsim 30$ degrees, while cosmological sGRBs above
the detection limit
appear to have narrower jet opening angles as a result of a correlation between $E_{\rm iso}$
and the beaming factor (e.g., Figure \ref{fig:beaming}). 

\begin{figure}[h!]
\centering
\includegraphics[width=\columnwidth]{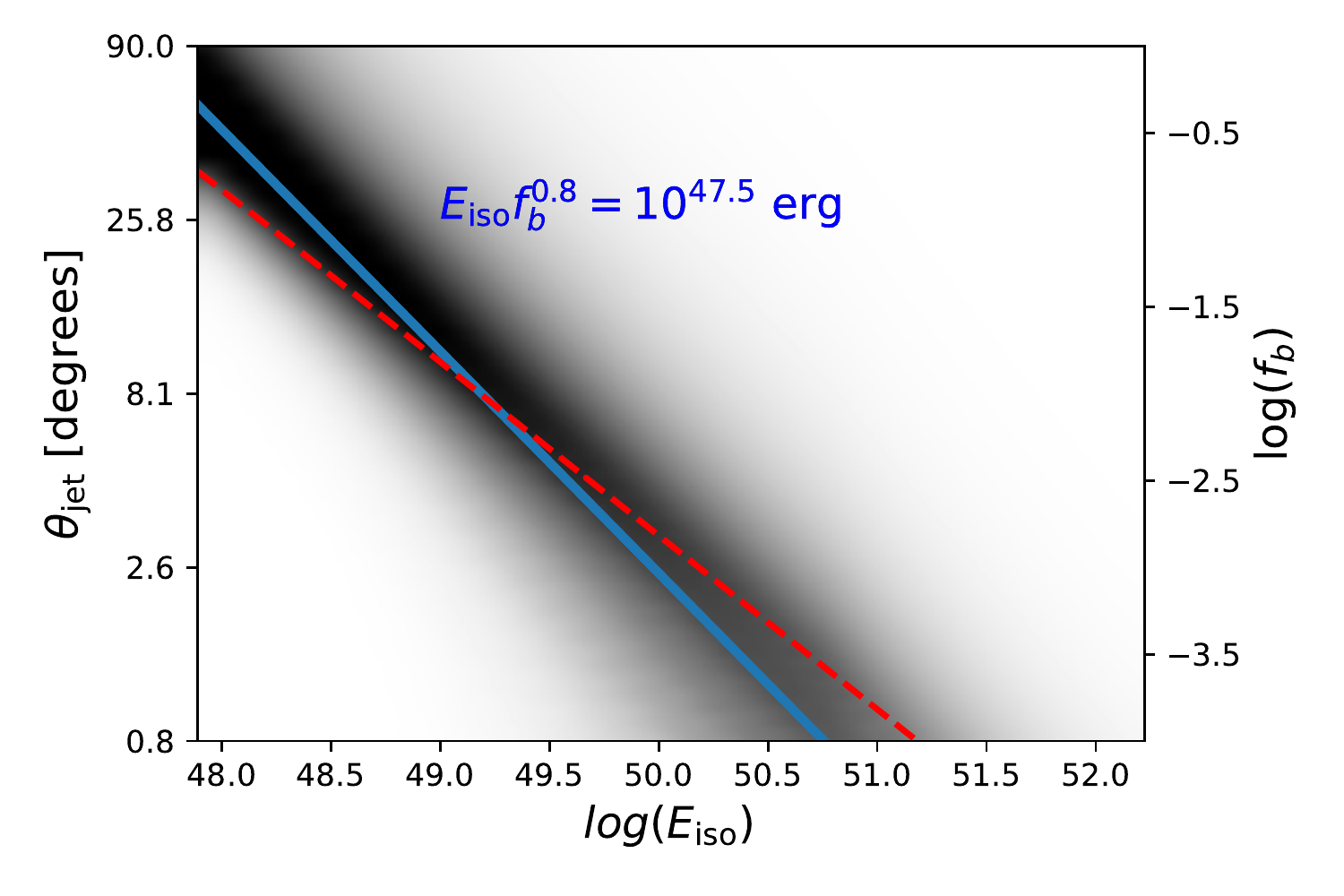}
\caption{The joint $E_{\rm iso}, f_b$ distribution at $z=0$. The
correlation between $E_{\rm iso}$ and $f_b$ appears as a
broad ridge around $E_{\rm iso} f_b^c = 10^{47.5}$ (solid blue curve) erg as a consequence 
of the steep luminosity function with a minimum energy (Section \ref{sec:obsdist}).
The similar curve corresponding to a ``standard'' sGRB $E_{\gamma}$ ($E_{\rm iso} \propto \theta_{\rm jet}^2$) using the small-angle formula is shown for reference (dashed
red curve).}
\label{fig:prob_eisofb}
\end{figure}

In the context of our sGRB world model (Section \ref{sec:models}), 
it is interesting to explore the uniqueness of GRB~170817A,
associated with GW170817 \citep{Abbott2017ApJ}.
The \textit{Fermi}/GBM fluence -- measured over the $T_{\rm 90}$ duration interval -- 
for GRB~170817A was $2.8 \times 10^{-7}$ erg cm$^{-2}$\citep[10-1000 keV;][]{Goldstein2017}.
The corresponding $E_{\rm iso} = 5.35 \times 10^{46}$ erg.
In our model (see, Figure \ref{fig:prob_eisofb}), such a low $E_{\rm iso}$
would be associated with a spherical, or nearly spherical explosion.
However, GRB~170817A was an exceptionally off-axis event, with an X-ray afterglow
that peaked 160 days after the sGRB \citep{Troja2020}.
Although the on-axis energetics remain unconstrained, it is likely
that the GRB was viewed at an angle $\theta_{\rm view} \sim 5\theta_{\rm jet}$ 
\citep[see,][and references therein]{Nakar2021ApJ}.
In our model, a broadly-jetted $\theta_{\rm jet}=15$ degree and $E_{\rm iso} = 10^{49}$ erg 
event is $10^2$--$10^3$ times more likely than a $\theta_{\rm jet}=5$ degree
event with energies comparable to those of the brightest, cosmological
sGRBs ($E_{\rm iso}\approx 10^{51}$--$10^{52}$ erg; Figure \ref{fig:obshisto}).

Although the off-axis nature of GRB~1701817A is clearly an
important clue as to the GW association, it does not suggest that we should increase our on-axis
rate predictions significantly toward our (larger) on- and off-axis rate predictions (Figure
\ref{fig:ligo}).
In the flat-top jet model \citep[e.g.,][]{Ioka2017} or
structured jet models \citep[e.g.,][]{Nakar2021ApJ}, the on-axis
event could be very bright ($E_{\rm iso}\gtrsim 10^{52}$ erg),
hence highly-unusual at such a nearby distance of 40 Mpc.
Similarly, if $E_{\rm iso} = 5.35 \times 10^{46}$ erg is instead taken to
be typical of the nearly-spherical emission that may accompany sGRBs, it is only
detectable at distances $D\lesssim 60$ Mpc by \textit{Swift}
or \textit{Fermi}, leading
to a predicted rate increase of only 16\% in a 200 Mpc aLIGO volume.

A Gaussian structured jet model can increase the observable solid angle significantly
for nearby events 
\citep[by a factor of 1.4 for $D<200$ Mpc, and 2.2 for $D<40$ Mpc; Figure \ref{fig:ligo}; see, also][]{Howell_2019}. 
Here, instead of $f_b = 1-\cos(\theta_{\rm jet})$, we write: 
\begin{equation}
f_b = \int \exp(-(\theta/\theta_{\circ})^2) \sin(\theta) d\theta.
\label{eq:structured}
\vspace{0.05in}
\end{equation}
To capture the same asymptotic behavior for small $\theta_{\rm jet}$ and
$\theta_{\rm jet}=90$ degrees, we set $\theta_{\circ}=\tan(\theta_{\rm jet})$.
Equation \ref{eq:structured} does not change the above analysis for cosmological GRBs which sample only the core of the jet; however the observable angles --
for nearby and narrowly-jetted events where the minimum-detectable $E_{\rm iso,min} < E_{\rm min}/f_b^c$ -- can be significantly wider than $\theta_{\rm jet}$. 
This is because the first factor of $f_b$ in Equation \ref{eq:r} is replaced with 
$1-\cos(\theta_{\rm view})$, where
$\theta_{\rm view}^2 = \theta_{\circ}^2\ln(L_{\gamma}/L_{\gamma,\rm min})$.
This also tends to favor narrowly-jetted events. For $D<40$ Mpc, the most likely $\theta_{\rm jet}=29$ degrees (see, Figure \ref{fig:beaming}) and the mean viewing angle is $\theta_{\rm view} \approx 60$ degrees.

The resulting sGRB flux is near the detection threshold. It is possible that such events
could be identified, likely in a non-decisive manner, after an optical counterpart is found and the number of search trials is reduced.

Predicted on-axis sGRBs within the aLIGO volume, however,
are well-above (factor of 26 on average) the \textit{Swift}/BAT
threshold. We estimate that the majority would be detected even using a down-scaled instrument with 80 times smaller effective
area. More important than sensitivity, 
\textit{Swift} itself can be expected to detect very few associations
(0.018 yr$^{-1}$ per 1.3 sr) due to it's narrow field of view. Experiments with large fields of
view, like \textit{Fermi}/GBM (0.09 yr$^{-1}$ per $2\pi$ sr), are far more likely to be successful.
To the extent that GRB~170817A is in any way typical, and off-axis emission is generally-detectable,
a sensitive and wide-field experiment like \textit{Fermi}/GBM is optimal 
(approaching 0.6 yr$^{-1}$ per $2\pi$ sr). 

Nonetheless, our estimates are fundamentally-limited by a scarcity of constraints on sGRB beaming. Additional sGRB detections and followup -- including deep followup
with Rubin Observatory \citep{Ivezic2019ApJ} to measure or limit the orphan afterglow \citep{Rhoads1999ApJ} rate -- is needed to better constrain sGRB jetting and to further probe the relation between sGRBs and GW events. 

Similarly, future GW detections in O4 and O5 with increasing sensitivity 
\citep[e.g.,][]{Williams_2018}, can further constrain the BNS rate beyond 200 Mpc 
and place more challenging constraints on sGRB beaming. Under the uncertainty in the fraction of GW events that result in jetting \citep{2021arXiv211103634T}, reproducing the GRB rate from the GW rate will be a greater challenge if not all GW events can produce a jet. If future GW observations constrain the BNS rate to $\lesssim 1$ Gpc$^{-1}$
(see, Section \ref{sec:params}), our starting assumption that all -- or even most --
sGRBs are due to compact object mergers is likely incorrect.

\bibliography{bib}
\bibliographystyle{aasjournal}

\end{document}